\documentclass[%
  reprint,
  superscriptaddress,
  amsmath,amssymb,
  pra,
  floatfix,
  longbibliography
]{revtex4-2}

\usepackage[T1]{fontenc}
\usepackage[utf8]{inputenc}
\usepackage{bm}
\usepackage{longtable}
\usepackage[colorlinks=true,allcolors=blue]{hyperref}
\usepackage{orcidlink}

\usepackage{braket}

\begin{document}

\title{Dirac's Dilemma of the Economy of Inheritance:\\
Parental Care, Equality of Opportunity, and Managed Inequality}

\author{Karl Svozil\,\orcidlink{0000-0001-6554-2802}}
\email{karl.svozil@tuwien.ac.at}
\homepage{http://tph.tuwien.ac.at/~svozil}

\affiliation{Institute for Theoretical Physics,
TU Wien,
Wiedner Hauptstrasse 8-10/136,
1040 Vienna,  Austria}

\date{\today}

\begin{abstract}
In a brief reflection on the principles of human society, P.~A.~M.~Dirac
articulated a structural tension between two widely affirmed norms:
that it is good and natural for parents to improve the prospects of their own
children, and that justice requires that all children have equal
opportunities in life.  These principles, each compelling on its own,
cannot be fully realized together.  This paper reconstructs Dirac's
dilemma, connects it to the dynamics of compounding advantage and
inheritance, and situates it within the broader history of political
philosophy, including the work of Rawls, Dworkin, Cohen, Brighouse and
Swift, Nozick, Murphy and Nagel, and others.  The paper argues that
attempts to eliminate the resulting injustices entirely risk damaging
the non--zero--sum structures that generate general prosperity, and
defends a position of ``managed inequality'': a robust social floor and
real mobility, combined with limits on extreme dynastic accumulation and
an explicit acceptance of some residual, but constrained, inherited
advantage.
\end{abstract}

\maketitle

\section{Introduction}

In a short talk later published under the title
``The Futility of War''~\cite{dirac-81},
P.~A.~M.~Dirac offers a strikingly simple statement of a deep social
problem.  He notes that modern human societies are guided by (at least)
two fundamental principles:
first, that it is good and natural for parents to care for and provide
for their children, even at substantial personal cost; and second, that
justice demands that all children be given an equal chance in life.
Dirac observes that these two principles, each attractive on its own,
stand in direct conflict once considered together.

In his own words (paraphrasing), if some parents manage to secure
better conditions for their children by making sacrifices, the children
of those parents will inevitably enjoy better opportunities than the
children of parents who do not or cannot do the same.  Conversely, if
the state really ensured that all children had equal chances, then there
would be no point in parents making such sacrifices, since their
children's prospects would be no better than those of others.  Dirac
concludes that these two principles ``tend to exclude each other'', and
that the conflict between them underlies familiar ideological divisions
between what he calls right--wing and left--wing outlooks.

Dirac does not present this as a solved problem; rather, he notes that
actual societies seek unstable compromises between these values, avoiding
extremes.  The aim of this paper is threefold.  First, to reconstruct
the structure of Dirac's dilemma.  Second, to connect it with the
dynamics of compounding advantage, both financial and non-financial.
Third, and centrally, to show how Dirac's brief remark fits into---and
is illuminated by---a much larger body of work in moral and political
philosophy concerning equality of opportunity, the family, inheritance,
and the tension between partiality and impartial justice.

The conflict that Dirac identifies is not unique to physics-minded
reflection.  It lies close to the surface of liberal egalitarian theory
(Rawls, Dworkin), of debates over ``luck egalitarianism'' and its
critics, of recent work on the ethics of the family (Brighouse \&
Swift; Cohen), of controversies about inheritance and property
(Nozick; Murphy \& Nagel), and of the long-standing philosophical worry
about balancing special obligations to intimates with impartial concern
for all (Sidgwick, Williams, Singer, Scheffler).  Seen from this
perspective, Dirac's contribution is a clear, compact formulation of a
problem that many philosophers have grappled with in more extended form.

\section{Dirac's formulation of the dilemma}

Dirac begins from two principles that he takes to be widely accepted in
modern societies:

\begin{enumerate}
  \item \textbf{Parental care and sacrifice.}  It is good, natural,
  and socially endorsed that parents care for their children, provide
  for them, and, where possible, improve their conditions of life, even
  at cost to themselves.  This principle, he notes, has clear analogues
  in the animal kingdom: many species invest heavily in their offspring.
  Human parents, similarly, often make substantial sacrifices to give
  their children better lives.

  \item \textbf{Equal opportunity for children.}  At the same time,
  Dirac argues, human beings possess a sense of justice that goes beyond
  animal behaviour.  It seems self-evident, he says, that all children
  born into the world ought to have an equal chance or equal
  opportunities for development, regardless of the circumstances of
  their birth.
\end{enumerate}

These two principles, however, are in immediate tension.  If some
parents manage, by effort and sacrifice, to provide their children with
better education, health, connections, or financial resources, then
those children will not have the same opportunities as children whose
parents either do not, or cannot, do the same.  The first principle
thus tends to produce inequalities in opportunities between children.

Conversely, if the second principle were realized strictly---if the
state really ensured that all children had equal chances independent of
parental behaviour---then additional parental sacrifice would not, in
fact, improve a child's prospects relative to others.  Parents could
still care for their children, but the space within which they could
\emph{alter} their relative chances would be heavily constrained.
Pursued to the limit, the second principle would thus erode the domain
within which the first operates effectively.

Dirac suggests that this structural conflict underlies the familiar
opposition between ``right--'' and ``left--wing'' ideologies.  In the
right--wing picture, parents are regarded as primarily responsible for
their children and are encouraged and praised when they make sacrifices
for them, even though this leads to unequal outcomes.  In the left--wing
picture, the state undertakes to secure equal opportunity for all
children and seeks to make that the only legitimate basis for
development, thereby limiting the role of parental advantage.  In
practice, he notes, actual countries adopt various compromises, seeking
to avoid the extremes of each position and the exacerbation of the
conflict they entail.

What Dirac does not do is offer a definitive resolution.  Rather, he
presents the clash of principles as a basic, perhaps permanent, problem
of human social life.

\section{Compounding inheritance and dynamic advantage}

Dirac's dilemma becomes sharper once we recognize that advantage is not
transmitted just once at birth but compounds over time and across
generations.  This compounding is both financial and non-financial.

\subsection{Financial compounding}

Consider a simple model of intergenerational wealth transmission.
Let $W_t$ denote the wealth of a family in generation $t$, and suppose
\begin{equation}
  W_{t+1} = (1 + r)\, W_t + Y_t,
\end{equation}
where $r > 0$ is an average rate of return on wealth, and $Y_t$
represents labour income in generation $t$.

Even if labour income $Y_t$ were equal across families---an assumption
far more egalitarian than reality---differences in initial wealth
$W_0$ would tend to grow over time via compounding.  Families that
begin with larger $W_0$ see their advantage expand roughly as
$(1+r)^t$, other things equal.  If, more realistically, higher wealth
also tends to generate better health, education, and access to lucrative
opportunities, thereby raising $Y_t$, the process is self-reinforcing.

This dynamic has been documented extensively in empirical work.  Thomas
Piketty, for example, argues that when the long-run rate of return on
capital $r$ exceeds the growth rate $g$ of the overall economy, wealth
accumulated in the past grows faster than output and income, leading to
increasing concentration of wealth and greater importance of
inheritance~\cite{Piketty2014}.  One need not accept all of Piketty's
conclusions to see that Dirac's conflict between parental
sacrifice and equal opportunity is not a one-off problem.  It is
amplified over time by compounding processes.

\subsection{Non-financial compounding}

Similar compounding occurs in non-financial domains.  Parents transmit
to their children various forms of \emph{human capital}
(skills, education, health), \emph{social capital} (networks, norms,
connections), and \emph{cultural capital} (language, expectations,
habits, dispositions).  These assets not only improve a child's position
at a given moment; they shape the entire trajectory of life, including
future earning potential and the ability to invest in the next
generation.

As a result, practices that instantiate Dirac's first principle---that
parents should strive to improve their children's prospects---create
patterns of advantage and disadvantage that, once established, tend to
persist and intensify.  Within such a dynamic, the second
principle---equal opportunity for all children---becomes harder, and
eventually impossible, to realize strictly.  The dilemma thus
acquires a historical and structural dimension.

\section{Dirac's dilemma in political philosophy}

While Dirac's articulation is compact and relatively late (the talk was
delivered in 1982), the underlying conflict between parental partiality
and egalitarian justice has been central to political philosophy since
at least the late nineteenth century.  In this section, I connect
Dirac's dilemma to several major strands of this literature.

\subsection{Rawls and fair equality of opportunity}

John Rawls's \emph{A Theory of Justice}~\cite{Rawls1999} is often taken as
the foundational work of contemporary liberal egalitarianism.  Rawls
famously defends two principles of justice, the second of which requires
both fair equality of opportunity and that social and economic
inequalities be arranged so that they are to the greatest benefit of the
least advantaged (the ``difference principle'').

Rawls's notion of \emph{fair equality of opportunity} goes substantially
beyond merely formal equality before the law.  It is not sufficient,
on his account, that offices and positions be \emph{formally} open to
all; rather, individuals with similar talents and willingness to use
them should have similar prospects of success, regardless of their
social background~\cite[sect.~12, 14]{Rawls1999}.  This is
very close to Dirac's second principle.

At the same time, Rawls insists that the family is part of the
``basic structure'' of society and thus a primary subject of justice.
He nonetheless recognizes that the existence of the family, as an
institution in which parents naturally seek to benefit their own
children, fundamentally constrains the extent to which fair equality of
opportunity can be realized.  He writes, in effect, that while
institutions should be arranged to mitigate the impact of social
contingencies on life chances, no feasible scheme of institutions can
eliminate all the effects of family background.\footnote{For explicit
discussion, see~\cite[sect.~17, 46]{Rawls1999}.  Rawls acknowledges
what he calls ``the tendency for the family to undermine the fair value
of the equal political liberties'' and the difficulty of fully
realizing fair equality of opportunity in the presence of the family.}

In other words, Rawls accepts something like Dirac's dilemma:
the family, and with it the first principle of parental care and
sacrifice, is both morally and practically indispensable, but its very
existence partly frustrates the realization of the egalitarian ideal of
fair equality of opportunity.  Rawls's response is, essentially, to
seek the best attainable approximation: a system of institutions that
significantly reduces, but does not and cannot eliminate, the unequal
effects of social starting points.

\subsection{Luck egalitarianism and the problem of the family}

Building on Rawls, a group of theorists sometimes called
``luck egalitarians'' sought to refine the ideal of equality by
distinguishing between inequalities that reflect individuals' voluntary
choices and those that reflect brute luck.  Roughly speaking, they
argue that justice requires neutralizing or compensating for
inequalities due to brute luck, while allowing inequalities that
result from responsible choices.

Ronald Dworkin's account of ``equality of resources'' is paradigmatic
here~\cite{Dworkin2000}.  Richard Arneson and G.~A.~Cohen develop
related views, arguing that inequalities in welfare or advantage are
only just when they trace to agents' choices rather than to
circumstances beyond their control~\cite{Arneson1989,Cohen2008}.

However, once one takes Dirac's first principle seriously, the family
poses acute problems for luck egalitarianism.  Many of the advantages
children enjoy reflect neither their choices nor their brute
luck, but the (permissible and valuable) partiality of their parents.
Parents choose to invest more or less in their children, and the
resulting inequalities in opportunity are not readily classifiable as
either deserved or undeserved by the children in luck-egalitarian
terms.

Critics such as Elizabeth Anderson have argued that this makes
luck egalitarianism morally unattractive.  In her influential paper
``What is the Point of Equality?'' she contends that luck egalitarian
policies would have to intrude intolerably into private life or condemn
people for unchosen features in ways that undermine their
self-respect~\cite{Anderson1999}.  Samuel Scheffler likewise emphasizes
the importance of ``associative duties'' to family and friends that
cannot be reduced to, or fully subordinated to, impartial
egalitarian norms~\cite{Scheffler2001}.

From Dirac's perspective, these debates can be seen as attempts to
respond to the very tension he identifies: strong egalitarian
commitments collide with the moral importance of family-based
partiality.  Luck egalitarians try to draw a principled line between
acceptable and unacceptable sources of inequality; their critics argue
that the line either fails to do justice to family life or leads to
objectionable consequences.

\subsection{The ethics of the family: Brighouse, Swift, and Cohen}

More recently, Harry Brighouse and Adam Swift have addressed Dirac's
dilemma almost directly in their book \emph{Family Values: The Ethics
of Parent--Child Relationships}~\cite{BrighouseSwift2014}.
They start from the observation that family relationships are
intrinsically valuable: they contribute to the flourishing of both
parents and children in ways that go beyond the distribution of
material resources.  At the same time, they recognize that allowing
parents to confer advantages on their children conflicts with the
ideal of equal opportunity.

Their central question is: \emph{How much inequality should we be
prepared to accept in order to sustain the goods associated with
family life?}  They argue that certain forms of parental partiality
(e.g.\ reading to one's children, spending time with them, providing
emotional support, modest gifts) are not only permissible but part of
what makes family life valuable.  However, they are more critical of
practices such as elite private schooling or large, untaxed
inheritances, which they regard as hard to justify when they
significantly undermine fair equality of opportunity.  Their proposed
solution is therefore a nuanced one: preserve the core goods of family
life while limiting those mechanisms through which parental advantage
translates into socially corrosive inequality.

G.~A.~Cohen addresses a related tension in his book
\emph{If You're an Egalitarian, How Come You're So Rich?}~\cite{Cohen2000}.
He argues that committed egalitarians who endorse strong principles of
distributive justice often live bourgeois lives that include extensive
familial partiality.  This, he suggests, reveals a kind of moral
incoherence: strict egalitarianism appears incompatible with many
ordinary, and seemingly admirable, family practices.  Cohen does not
resolve the conflict so much as expose its depth and urge greater
reflective honesty about it.

Both Brighouse \& Swift and Cohen thus deepen Dirac's intuition:
justice and parental partiality are not easily harmonized.  The
question is not how to fully reconcile them, but how much injustice we
are willing to tolerate for the sake of valuable personal
relationships, and how far we are prepared to constrain those
relationships for the sake of justice.

\subsection{Inheritance, property, and the legitimacy of bequest}

Dirac's second principle---equal chances for all children---also
connects to older debates about inheritance, property rights, and
redistribution.

On one side, libertarian theorists such as Robert Nozick defend a
strong right to private property and voluntary transfer, including
bequest~\cite{Nozick1974}.  In his ``entitlement theory'', a
distribution of holdings is just if everyone is entitled to the
holdings they possess under the principles of justice in acquisition
and transfer, regardless of how unequal the outcome may be.  On this
view, there is no independent requirement that distributions be
egalitarian or that inequalities be justified by their effects on the
worst off.  Parents are entitled to dispose of their property, including
by leaving it to their children, and the resulting inequalities in
opportunity are morally acceptable if they arise from just processes.

On the other side, egalitarians have argued that strong inheritance
rights are hard to reconcile with a commitment to equality of
opportunity.  Liam Murphy and Thomas Nagel, for example, emphasize that
property rights themselves are defined and sustained by the legal and
tax systems, and that claims about what individuals ``deserve'' cannot
be detached from the background institutions that structure economic
life~\cite{MurphyNagel2002}.  From their perspective, substantial
taxation of large inheritances can be justified as part of a fair
overall system, especially when such inheritances contribute to
unacceptable inequalities in life chances.

Economists like Piketty add empirical weight to these concerns,
arguing that, absent progressive taxation and other checks, modern
economies tend to evolve toward forms of ``patrimonial capitalism'' in
which inherited wealth once again dominates individual
achievement~\cite{Piketty2014}.  This is effectively a large-scale,
data-driven illustration of the compounding process that underlies
Dirac's worry.

These debates show that Dirac's dilemma is not merely about the
\emph{attitudes} of parents and states, but about the design of concrete
institutions governing property, taxation, education, and family law.
Different positions in the inheritance debate correspond to different
implicit resolutions of the conflict between parental freedom to confer
advantage and the egalitarian aspiration to equalize children's life
chances.

\subsection{Partiality and impartial morality}

At a more abstract level, Dirac's tension exemplifies a familiar and
wider problem: the conflict between impartial moral principles that
treat all persons' interests as equally important, and partial
obligations to family, friends, and associates.

Henry Sidgwick, in \emph{The Methods of Ethics}, already notes the
``dualism of practical reason'' between egoistic and utilitarian
reasons, and more generally the tension between special obligations and
impartial benevolence~\cite{Sidgwick1907}.  Later thinkers such as
Bernard Williams stress that moral theories which demand pervasive
impartiality can alienate individuals from their own projects and
relationships~\cite{Williams1981}.  Peter Singer, in contrast,
famously argues that we are morally required to make substantial
sacrifices to help distant strangers, a view that, if taken seriously,
leaves much less moral room for privileging our own
children~\cite{Singer1972}.

Samuel Scheffler, in \emph{Boundaries and Allegiances}, articulates a
middle position: he maintains that special responsibilities to family
and associates are genuine and morally weighty, but also that they are
subject to constraints of justice and cannot entirely override
impartial considerations~\cite{Scheffler2001}.  This is very much in
the spirit of Dirac's own suggestion that societies must strike a
pragmatic compromise between conflicting but individually compelling
principles.

\section{Non--zero--sum considerations and the cost of radical equalization}

The intuitive response to the compounding of inherited advantage is to
push more firmly for equalization: if uncorrected parental partiality
generates severe inequalities in opportunity, perhaps the state should
aggressively redistribute resources and constrain the ways in which
parents can favour their own children.  Dirac himself gestures toward
this thought when he associates the second principle with left--wing
ideology and the first with right--wing ideology.

However, two facts are crucial here:

\begin{enumerate}
  \item Modern economic and social life is not a pure zero--sum game.
  \item People respond behaviourally to institutions; incentives shape
  investment, effort, and innovation.
\end{enumerate}

If total wealth and productive capacity were fixed, questions of justice
would concern only how to divide a given pie.  In that case, correcting
inequalities from inheritance would, at least in principle, leave total
resources unchanged.  In reality, institutions and policies influence
not only the \emph{distribution} but also the \emph{size} of the total
product.  The ways in which property rights, taxation, and family
policies are designed affect whether people invest, save, and take
productive risks.

Some redistributive measures are compatible with, and may even promote,
long-run prosperity: for example, public investment in education and
health, or moderate progressive taxation to fund an effective safety
net.  But beyond a certain point, especially when expropriation or
confiscatory taxation threatens the security of holdings, individuals
may reduce investment, withdraw effort, or move their activities into
informal or illicit channels.  Historical episodes of radical
levelling---such as aspects of the early Soviet revolution---have
sometimes led not to a stable egalitarian order, but to economic
collapse, repression, and generalized scarcity.

From a Rawlsian perspective, this non--zero--sum character of social
cooperation is reflected in the difference principle: inequalities in
income and wealth are just only if they work to the advantage of the
least advantaged over time~\cite{Rawls1999}.  Eliminating all
inheritance and sharply curtailing parental partiality might move
society closer to strict equality of opportunity, but if doing so
substantially reduces the overall level of wealth and opportunity, the
position of the worst off may, in fact, deteriorate.

Dirac's dilemma thus interacts with a second tension: between the
\emph{depth} of egalitarian correction one seeks and the \emph{capacity}
of the social--economic system to generate opportunity in the first
place.  Ignoring the non--zero--sum structure risks policies that
destroy the conditions for the very justice they aim to secure.

To this theoretical concern, the historian Walter Scheidel adds a sobering empirical dimension.
In \emph{The Great Leveler}~\cite{Scheidel-2017}, Scheidel argues that throughout human history,
substantial reductions in inequality have almost exclusively resulted from violent shocks---what he terms
the ``Four Horsemen'': mass mobilization warfare, transformative revolution, state failure, and lethal pandemics.
Peaceful mechanisms such as increased taxation, land reform or democratization have historically failed to generate comparable leveling effects.
 This suggests that the tension identified by Dirac is not merely theoretical but historically stubborn:
attempts to strictly realize the second principle have typically required the catastrophic destruction
of wealth rather than its `` soft'' redistribution.

\section{Towards a normative position: floors, mobility, and managed inequality}

Given the persistent conflict between family-based partiality and
egalitarian justice, and the non--zero--sum character of modern
economies, what kind of institutional response is plausible?

\subsection{From strict equality to robust opportunity}

One important step is to shift the focus from strict equality of
opportunity to a combination of:
\begin{enumerate}
  \item a robust \emph{floor} of opportunities and life conditions for
  every child, and
  \item genuine \emph{social mobility}, such that talent and effort can
  make a substantial difference to life prospects.
\end{enumerate}

A robust floor includes ensuring that every child has access to:
\begin{itemize}
  \item adequate nutrition, healthcare, and early childhood support;
  \item safe and stable housing;
  \item effective basic education and the possibility of further
  education or training;
  \item protection from abuse, neglect, and extreme deprivation.
\end{itemize}

These are not marginal goods; they require significant public resources
and institutional capacity.  Yet they still fall short of eliminating
all differences in parental endowments and investments.  Some children
will receive more extensive tutoring, richer cultural experiences, and
stronger social networks from their parents than others.  Under this
model, justice does not demand the erasure of all such differences, but
insists that no child be relegated to a life of severely stunted
opportunity due to circumstances of birth.

This approach aligns with Rawls's emphasis on ensuring both fair
equality of opportunity (to the extent feasible) and a social minimum
for the least advantaged, while recognizing, as Rawls himself does,
that the family will always limit how far equality of opportunity can
go~\cite{Rawls1999}.  It also resonates with Brighouse and Swift's
effort to protect the core goods of family life while curbing the most
harmful forms of educational and economic advantage.

\subsection{Limiting runaway dynastic accumulation}

Beyond establishing a floor and promoting mobility, societies can
legitimately act to prevent extreme and self-reinforcing concentrations
of wealth and power.  The aim here is not to render initial conditions
identical, but to avoid a situation in which birth overwhelmingly
determines prospects and a quasi-hereditary class structure emerges.

In institutional terms, this might involve:
\begin{itemize}
  \item moderate but effective taxation of large inheritances and
  estates, with thresholds high enough not to penalize modest
  intergenerational transfers;
  \item policies that target \emph{rent-seeking} forms of wealth
  (monopolies, regulatory capture, unearned privileges) more intensively
  than wealth generated by innovation and productive investment;
  \item strong institutions that protect open competition, the rule of
  law, and merit-based access to education and employment;
  \item regulation of elite educational and professional ``gates''
  whose closure would otherwise allow advantage to harden into
  caste-like structures.
\end{itemize}

These measures accept that some inherited advantage is both inevitable
and, within limits, acceptable, but seek to prevent the kind of
compounding dynastic dominance that Dirac, Piketty, and many
egalitarians find troubling.

\subsection{Accepting managed injustice}

From Dirac's starting point, as enriched by the philosophical literature
reviewed above, the central normative conclusion is uncomfortable but
difficult to avoid: some degree of injustice---in the sense of
deviation from strict equality of opportunity---is not merely tolerated
but instrumentally justified.  The reason is not that inequality is
intrinsically valuable, but that the attempt to eliminate it entirely
would so disrupt the economic and social conditions of cooperation, and
so intrude into vital personal relationships, that overall
well-being---including that of the worst off---would plausibly be lower.

This position of ``managed inequality'' is close in spirit to Rawls's
difference principle, to Brighouse and Swift's defence of family
values within limits, and to Scheffler's account of associative duties
constrained by justice.  It rejects both the radical
levelling that would extinguish much of the space for parental
partiality and the laissez-faire acceptance of compounding hereditary
advantage that effectively abandons the egalitarian aspiration.

The task of political design, on this view, is not to resolve Dirac's
dilemma once and for all, but to structure institutions so that:
\begin{itemize}
  \item parental care and sacrifice can flourish within a protected
  domain;
  \item no child falls below a decent social minimum or is
  systematically excluded from meaningful opportunities;
  \item extreme concentrations of wealth and privilege are checked
  before they ossify into caste-like structures;
  \item the productive and innovative capacities of the society are
  sustained and directed in ways that benefit the least advantaged.
\end{itemize}

\section{Conclusion}

Dirac's brief reflection on the futility of war contains a penetrating
diagnosis of a basic problem of human society: the structural conflict
between the natural and morally valued impulse of parents to advance
their own children's interests, and the equally compelling ideal that
all children should have equal opportunities in life.  This conflict is
amplified over time by compounding financial and non-financial
advantages and is intertwined with the non--zero--sum character of
modern social cooperation.

The history of political philosophy confirms that there is no simple
reconciliation of these principles.  Rawls, Dworkin, Cohen, Brighouse
\& Swift, Nozick, Murphy \& Nagel, Piketty, and many others have, in
different ways, grappled with the tension between family-based
partiality, equality of opportunity, and the institutional regulation of
inheritance and wealth.  The upshot is not a tidy solution, but a
recognition that some trade-off is unavoidable: we must decide, as a
matter of principle and policy, how much inequality we are willing to
accept for the sake of parental freedom and how far we are prepared to
limit family-based advantage in the name of justice.

The position sketched here favours a deliberately managed compromise:
a robust social floor and real mobility, sustained by institutions that
foster both opportunity and productivity, combined with controls on
runaway dynastic accumulation.  Within that framework, a bounded amount
of inherited advantage and resulting inequality is accepted, not because
it is just by itself, but because efforts to eradicate it entirely would
destroy the conditions for a just and flourishing society.

In this sense, Dirac's dilemma is not a problem to be solved and set
aside, but a permanent feature of moral and political life, calling for
continuous reflection, adjustment, and restraint.

\bibliography{svozil}

\end{document}